\documentclass{elsart1p}
\usepackage{graphics}
\usepackage{graphicx}
\usepackage{epsfig}
\usepackage{amssymb}

\usepackage[normalem]{ulem}  
\usepackage{color} 

\renewcommand\sout{\bgroup \color{red} \ULdepth=-.5ex \ULset}

\begin{document}

\begin{frontmatter}

\title{Nuclear thermodynamics and the \\in-medium chiral condensate}
\author[]{Salvatore Fiorilla},
\author[]{Norbert Kaiser} \author[]{and Wolfram Weise}
\address{Physik-Department, Technische Universit\"at M\"unchen, D-85747 
Garching, Germany}

\begin{abstract}
The temperature dependence of the chiral condensate in isospin-symmetric 
nuclear matter at varying baryon density is investigated using thermal 
in-medium chiral effective field theory. This framework provides a realistic 
approach to the thermodynamics of the correlated nuclear many-body system and 
permits calculating systematically the pion-mass dependence of the free 
energy per particle. One- and two-pion exchange processes, 
$\Delta(1232)$-isobar excitations, Pauli blocking corrections and three-body 
correlations are treated up to and including three loops in the expansion of 
the free energy density. It is found that nuclear matter remains in the 
Nambu-Goldstone phase with spontaneously broken chiral symmetry in the 
temperature range $T\lesssim 100\,$MeV and at baryon densities at least up to 
about twice the density of normal nuclear matter, $2\rho_0 \simeq 0.3\,
$fm$^{-3}$. Effects of the nuclear liquid-gas phase transition on the chiral 
condensate at low temperatures are also discussed.
\end{abstract}

\end{frontmatter}


The chiral condensate $\langle \bar{q}q \rangle$, i.e. the expectation value 
of the scalar quark density, plays a fundamental role as an order parameter 
of spontaneously broken chiral symmetry in the hadronic low-energy phase of 
QCD. The variation of $\langle \bar{q}q \rangle$ with temperature and baryon 
density is a key issue for locating the chiral transition boundary in the QCD 
phase diagram. The melting of the condensate at high temperatures and/or
densities determines the crossover from the Nambu-Goldstone phase to the 
Wigner-Weyl realization of chiral symmetry in QCD. 

It is thus of principal interest to perform a systematically organized
calculation of the thermodynamics of the chiral condensate. Such a 
calculation requires knowledge of the dependence of the free energy density on 
the light quark mass (or equivalently, on the pion mass). The 
appropriate framework for such a task is in-medium chiral effective field 
theory with its explicit access to one- and two-pion exchange dynamics and 
resulting two- and three-body correlations in the presence of a nuclear 
medium. 

Previous studies of the in-medium variation of the chiral condensate were
mostly concerned with the density dependence of  $\langle \bar{q}q \rangle$ 
at zero temperature, using different approaches such as QCD sum rules 
\cite{drukarev} or models \cite{LiKo94,bw96} based on the boson exchange 
phenomenology of nuclear forces. Temperature effects have been included in 
schematic Nambu-Jona-Lasinio (NJL) approaches \cite{HK,KLW}. Such
NJL models work with quarks as quasiparticles and provide useful insights 
into dynamical mechanisms behind spontaneous chiral symmetry breaking and 
restoration, but  they do not properly account for nucleons and their 
correlations, a prerequisite for a more realistic treatment.    

The present work extends a previous chiral effective field theory calculation 
\cite{cond} of the density-dependent in-medium condensate $\langle \bar{q}q 
\rangle(\rho)$ to finite temperatures $T$. Corrections to the linear density 
approximation are obtained by differentiating the interaction parts of the 
free energy density of isospin-symmetric nuclear matter with respect to the 
(squared) pion mass. Effects from one-pion exchange (with $m_\pi$-dependent 
vertex corrections), iterated $1\pi$-exchange, and irreducible $2\pi$-exchange 
including intermediate $\Delta(1232)$-isobar excitations, with Pauli-blocking 
corrections are systematically treated up to three-loop order. The dominant 
nuclear matter effects on the dropping condensate are supplemented by a 
further small reduction due to interacting thermal pions. To anticipate the 
result: we find that the delayed tendency towards chiral symmetry restoration 
with increasing baryon density $\rho$, observed at $T=0$ \cite{cond} in the 
same framework, gets gradually softened with increasing temperature. An 
approximately linear decrease of the quark condensate with increasing $\rho$ 
is recovered at temperatures around $T\simeq 100\,$MeV. However, no rapid drive
towards a first order chiral phase transition is seen, at least up to $\rho 
\lesssim 2\,\rho_0$ where $\rho_0 = 0.16$ fm$^{-3}$ is the density of normal 
nuclear matter.  The only phase transition known for nuclear matter in this 
density range is the first order transition between an interacting Fermi gas 
and a Fermi liquid, with its broad coexistence region extending from low 
densities up to about $\rho_0$. Signatures of this liquid-gas transition are 
nonetheless visible also in the chiral condensate at low temperatures and 
will be discussed in this work. 

Our starting point is the free energy density, ${\cal F}(\rho,T) = \rho \bar 
F(\rho,T)$, of isospin-symmetric spin-saturated nuclear matter, with 
$\bar{F}(\rho,T)$ the free 
energy per particle. In the approach to nuclear matter based on in-medium 
chiral perturbation theory \cite{nucmatt,deltamat,FKW2012}  the free energy 
density is given by a sum of convolution integrals of the form, 
\begin{eqnarray} \rho \, \bar F(\rho,T)&=& 4\int_0^\infty dp\, p \, {\cal K}_1
\,n(p)+\int_0^\infty dp_1\int_0^\infty dp_2\, {\cal K}_2\, n(p_1)
n(p_2)\nonumber \\ &+& \int_0^\infty dp_1\int_0^\infty dp_2\int_0^\infty dp_3\, 
{\cal K}_3 \,n(p_1)n(p_2)n(p_3)+\rho \, \bar{\cal A}(\rho,T)\,, \end{eqnarray}
where ${\cal K}_1, {\cal K}_2$ and ${\cal K}_3$ are one-body, two-body and 
three-body kernels, respectively. The last term, the so-called anomalous 
contribution $\bar {\cal A}(\rho,T)$ is a special feature at finite 
temperatures \cite{kohn} with no counterpart in the calculation of 
the ground state energy density at $T=0$. As shown in ref.\cite{nucmatt} the 
anomalous contribution arising in the present context from second-order 
pion exchange has actually very little influence on the equation of state of 
nuclear matter at moderate temperatures $T< 50\,$MeV. 

The quantity  
\begin{equation} 
n(p) = {p\over 2\pi^2} \bigg[ 1+\exp{p^2/2M_N -\tilde \mu \over T} \bigg]^{-1} 
\end{equation}
denotes the density of nucleon states in momentum space. It is the product of 
the temperature dependent Fermi-Dirac distribution and a kinematical prefactor 
$p/ 2\pi^2$ which has been included in $n(p)$  for convenience. 
$M_N$ stands for the (free) nucleon mass. The particle density $\rho$ is 
calculated as
\begin{equation}\rho= 4\int_0^\infty dp\, p \,n(p) \,. \end{equation} 
This relation determines the dependence of the effective one-body 
chemical potential $\tilde \mu(\rho,T;M_N)$ on the thermodynamical 
variables $(\rho, T)$ and indirectly also on the nucleon mass $M_N$. 
The one-body kernel ${\cal K}_1$ in eq.(1) provides the contribution of the
non-interacting nucleon gas to the free energy density and it reads
\cite{nucmatt}:
\begin{equation} \label{K1}
{\cal K}_1(p) = M_N +\tilde \mu- {p^2\over 3M_N}- {p^4\over 8M_N^3} \,. 
\end{equation}
The first term in ${\cal K}_1$ gives the leading contribution (density 
times nucleon rest mass) to the free energy density. The 
remaining terms account for (relativistically improved) kinetic energy 
corrections.

The two- and three-body kernels, ${\cal K}_2$ and ${\cal K}_3$, specifying 
all one- and two-pion exchange processes up to three loop order for the
free energy density, have already been given in explicit form in 
refs.\cite{nucmatt,deltamat,FKW2012} and will not be repeated here. We recall 
from our earlier works that after fixing only a few contact terms the 
free energy density computed from these interaction kernels provides a 
realistic nuclear equation of state up to densities $\rho \lesssim 2\,\rho_0$.
 What matters in the following will be the dependence of the kernels 
${\cal K}_2$ and ${\cal K}_3$ on the light quark mass, $m_q$, or equivalently, 
on the pion mass, $m_\pi$, that is introduced by pion propagators and by pion 
loops. In-medium chiral effective field theory is the appropriate framework 
to quantity this pion-mass dependence in a systematic and reliable way.
 
Application of the Feynman-Hellmann theorem  establishes an exact connection 
between the temperature and density dependent in-medium quark condensate 
$\langle \bar q q\rangle(\rho,T)$ and the derivative of the free energy 
density of (isospin-symmetric, spin-saturated) nuclear matter with respect to 
the light quark mass $m_q$. Using the Gell-Mann-Oakes-Renner relation $m_\pi^2 
f_\pi^2 = -m_q \langle 0|\bar q q|0\rangle$, one finds for the ratio of the 
in-medium to vacuum quark condensate
\begin{equation}  
{\langle \bar q q\rangle(\rho,T)\over  \langle 0|\bar q q|0
\rangle} = 1 - {\rho \over f_\pi^2} {\partial \bar F(\rho,T) \over \partial 
m_\pi^2} \,,\end{equation}
where the derivative with respect to $m_\pi^2$ is to be taken at fixed $\rho$ 
and $T$. The quantities  $\langle 0|\bar q q|0\rangle $ (vacuum quark 
condensate) and $f_\pi$ (pion decay constant) are to be understood as 
taken in the chiral limit, $m_q\to 0$. Likewise, $m_\pi^2$ stands for the 
leading linear term in the quark mass expansion of the squared pion mass. 

In the one-body kernel  ${\cal K}_1$ the quark (or pion) mass 
dependence is implicit via its dependence on the nucleon mass $M_N$. The 
condition $\partial \rho/\partial M_N=0$ applied to eq.(3) leads to the 
following dependence of the effective one-body chemical potential $\tilde \mu$ 
on the nucleon mass $M_N$:
\begin{equation} {\partial \tilde \mu \over \partial M_N }= {3 \rho \over 2M_N 
\Omega_0''} \,, \qquad   \Omega_0''= -4M_N  \int_0^\infty dp\, 
{n(p) \over p}\,. \end{equation} 
The nucleon sigma term $\sigma_N = \langle N|m_q \bar q q |N\rangle= m_\pi^2 \,
\partial M_N/\partial m_\pi^2$ measures the variation of the nucleon mass with
the quark (or pion) mass. Combining both relationships leads to the 
following $m_\pi^2$-derivative of the one-body kernel:
\begin{equation} {\partial {\cal K}_1 \over \partial m_\pi^2} = {\sigma_N 
\over m_\pi^2} \bigg\{ 1+ {3 \rho \over 2M_N \Omega_0''} +{p^2 \over 3M_N^2} 
+{3p^4 \over 8M_N^4} \bigg\}\,. \end{equation}
In the limit of zero temperature, $T=0$, the terms in eq.(7) reproduce the 
linear decrease of the chiral condensate with density. The kinetic 
energy corrections account for the (small) difference between the scalar and 
the vector (i.e. baryon number) density. In the actual calculation we  
use the chiral expansion of the nucleon sigma term $\sigma_N$ to order 
${\cal O}(m_\pi^4)$ as given in eq.(19) of ref.\cite{cond}. The empirical 
value of the nucleon sigma term (at the physical pion mass $m_\pi = 135\,$MeV) 
is $\sigma_N =(45\pm 8)\,$MeV \cite{gls}. Recent results for the quark mass 
dependence of baryon masses from lattice QCD and accurate chiral 
extrapolations \cite{sigmaterm} tend to give smaller values of this sigma 
term, but still consistent with the empirical $\sigma_N$ within errors. We 
use the central value $\sigma_N = 45$\,MeV in the present calculations,
keeping in mind however that the existing error band in the determination of 
this quantity is still a primary source of uncertainties in following 
discussion.

For the two- and three-body kernels, ${\cal K}_2$ and ${\cal K}_3$, 
related to one-pion exchange and iterated one-pion exchange, explicit 
expressions have been given in ref.\cite{nucmatt}. Their derivatives with 
respect to the squared pion mass, $\partial {\cal K}_{2,3}/\partial m_\pi^2$, 
are hence obvious  and do not need to be written out here. The same applies 
to the anomalous contribution $\bar {\cal A}(\rho,T)$ (see eqs.(14,15) in 
ref.\cite{nucmatt}) and to the two- and three-body kernels related to 
$2\pi$-exchange with excitation of virtual $\Delta(1232)$-isobars (see 
section 6 in ref.\cite{deltamat}). In case of the one-pion exchange 
contribution we include the $m_\pi$-dependent vertex correction factor 
$\Gamma(m_\pi)$ as discussed in section 2.1 of ref.\cite{cond}. The 
short-distance contact term that produces a $T$-independent correction of 
order $\rho^2$ to the in-medium condensate is treated  exactly in the same 
way as in ref.\cite{cond}, i.e. all terms with a non-analytical quark-mass 
dependence generated by pion-loops are taken into account.  

\begin{figure}[htb]\begin{center}
\includegraphics[width=5.5cm]{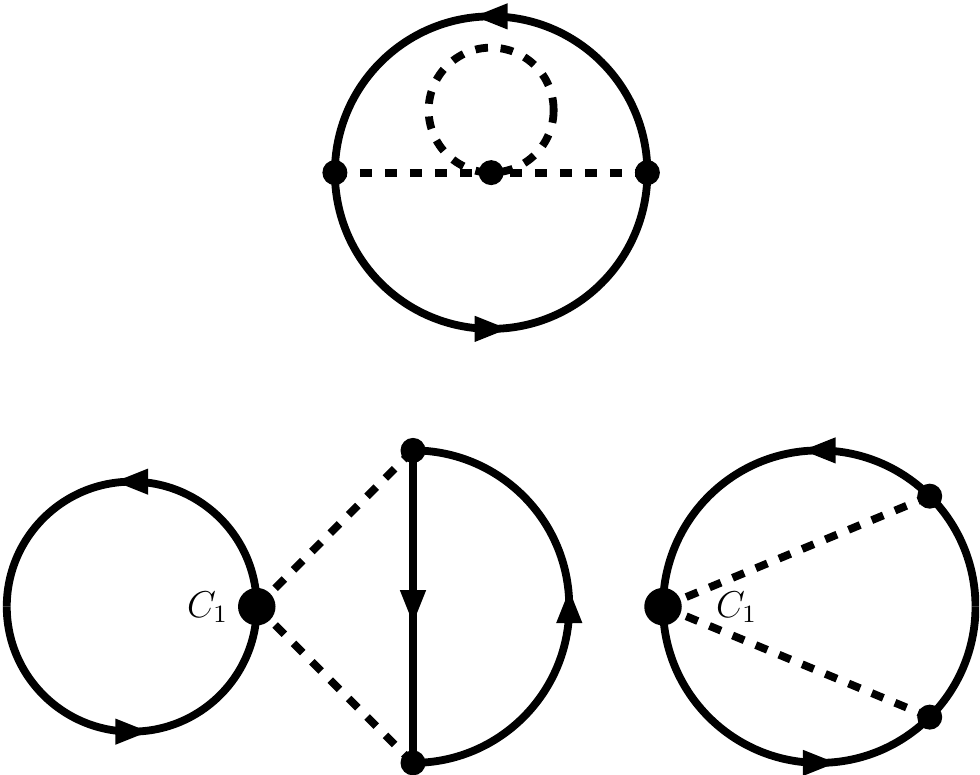}
\end{center}
\caption{Three-loop contributions to the free energy density of nuclear 
matter that are relevant for the in-medium chiral condensate. Upper diagram:
pion self-energy correction; lower diagrams: two-pion exchange Hartree and Fock
terms involving the $\pi\pi NN$ contact interaction proportional to the 
low-energy constant $c_1$.}
\label{figure1}\end{figure}
\begin{figure}[htb]\begin{center}
\includegraphics[width=9.8cm]{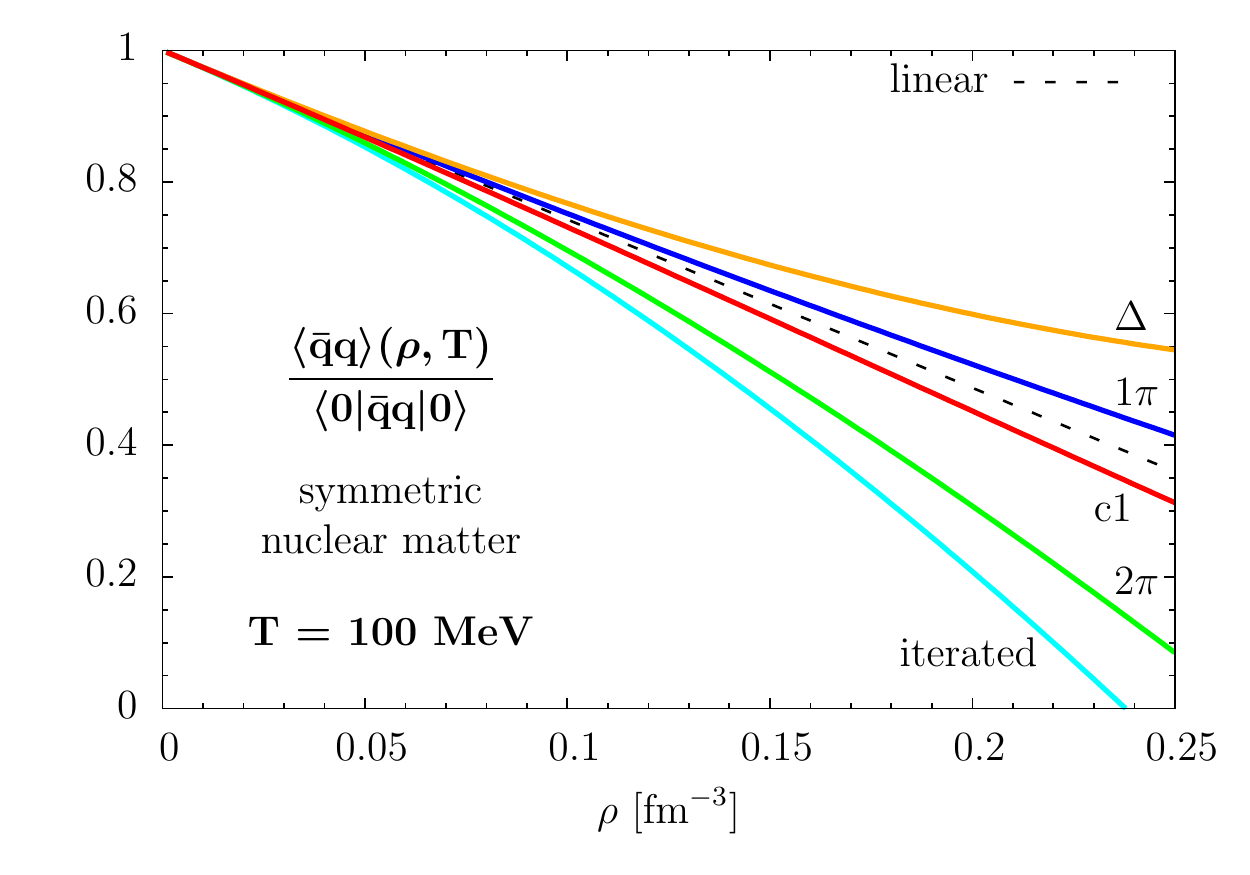}
\end{center}
\caption{Density dependence of the chiral condensate in isospin-symmetric 
nuclear matter at temperature $T = 100$ MeV. Starting from the linear density 
dependence (dashed curve) characteristic of the free nucleon Fermi gas, the 
following interaction contributions are successively added: one-pion exchange 
Fock term ($1\pi$), second order (iterated) pion exchange, irreducible 
two-pion exchange ($2\pi$), two- and three-body contributions from 
$2\pi$ exchange with intermediate $\Delta$ excitations ($\Delta$), and 
two-pion exchange with $\pi\pi NN$ vertex involving the low-energy constant 
$c_1$. Pauli blocking effects are included throughout.}
\label{figure2}
\end{figure}

Let us now turn to some three-loop contributions that are new and of special 
relevance for the in-medium chiral condensate. The first one comes from the 
pion self-energy diagram shown in Fig.1. It gives rise to the following 
$m_\pi^2$-derivative of the two-body kernel:
\begin{eqnarray} 
{\partial {\cal K}_2^{(\pi)}\over \partial m_\pi^2 }
&=& {3g_A^2 m_\pi^2 \over \pi^2 (4f_\pi)^4} \Big\{ \bar{\ell}_3\big[(X^-_{12})^2 -
(X^+_{12})^2\big] \nonumber\\ && + (4\bar{\ell}_3 - 1)\left(X^+_{12} - 
X^-_{12} \right)+ (2\bar{\ell}_3 - 1)\ln{X^-_{12}\over X^+_{12}}\Big\}\,,
\end{eqnarray}
with the abbreviations $X^\pm_{ij} = [1 + (p_i\pm p_j)^2/m_\pi^2]^{-1}$ and the 
$\pi\pi$ low-energy constant $\bar{\ell}_3  \simeq 3$. 

The chiral $\pi\pi NN$ contact vertex proportional to $c_1 m_\pi^2 $ generates 
$2\pi$-exchange Hartree and Fock diagrams, also shown in Fig.1. Concerning the 
free energy density $\rho \bar F(\rho,T)$ or the equation of state of nuclear 
matter their contributions are actually almost negligible. However,
when taking the derivative with respect to $m_\pi^2$ as required for the 
calculation of the in-medium condensate, these contributions turn out to be 
of  similar importance as other interaction terms. The corresponding 
contribution to the derivative of 
the two-body kernel reads:
\begin{equation} {\partial {\cal K}_2^{(c_1)}\over \partial m_\pi^2 } = {g_A^2 
c_1 m_\pi^3 \over 8\pi f_\pi^4} \bigg\{ G\Big({p_1+p_2 \over 2m_\pi} \Big)- 
 G\Big({p_1-p_2 \over 2m_\pi} \Big) \bigg\}\,,  \end{equation}
with the auxiliary function:
\begin{equation} G(x)=8x(3+x^2)\arctan x -5\ln(1+x^2)-100 x^2\,.\end{equation}
The $2\pi$-exchange Hartree diagram with one $c_1m_\pi^2$-vertex contributes
the following piece to the three-body kernel: 
\begin{equation}  
{\partial {\cal K}_3 ^{(c_1,H)}\over \partial m_\pi^2}=  {6g_A^2 c_1 p_3\over 
f_\pi^4} \Big\{(X^+_{12}-X^-_{12})(X^+_{12}+X^-_{12}-3) +\ln {X^+_{12}\over X^-_{12}} 
\Big\}\,,\end{equation}
while the three-body term associated with the  $2\pi$-exchange Fock
diagram with one $c_1m_\pi^2$-vertex gives:
\begin{eqnarray} {\partial {\cal K}_3^{(c_1)}\over \partial m_\pi^2 } 
&=&  {3g_A^2 c_1  \over f_\pi^4} \bigg[{p_2\over p_3} +{p_3^2-p_2^2-m_\pi^2 \over 
4p_3^2} \ln {X^-_{23}\over X^+_{23}}  \bigg] \nonumber\\ && \times \bigg[
p_1+{p_3^2-p_1^2-3m_\pi^2 \over 4p_3} \ln {X^-_{13} \over X^+_{13}}
+(p_1+p_3)X^+_{13} +(p_1-p_3) X^-_{13}\, \bigg]\,. \end{eqnarray}
Last not least we incorporate the effects of thermal pions. Through its  
$m_\pi^2$-derivative the pressure (or free energy density) of 
thermal pions gives rise to a further reduction of the $T$-dependent 
in-medium condensate. In the two-loop approximation of chiral perturbation 
theory including effects from the $\pi\pi$-interaction one finds the following
shift of the condensate ratio in the presence of the pionic heat bath 
\cite{gerber,toublan,pipit}: 
\begin{eqnarray}
{\delta\langle \bar q q\rangle(T)\over  \langle 0|\bar q q|0
\rangle} &=&-{3m_\pi^2 \over (2\pi f_\pi)^2} H_3\Big({m_\pi 
\over T}\Big) \bigg\{1+ {m_\pi^2 \over 8\pi^2 f_\pi^2} 
\bigg[H_3\Big({m_\pi \over T}\Big)-  H_1\Big({m_\pi \over T}\Big) + {2-3
\bar{\ell}_3 \over 8} \bigg] \bigg\}\,, \nonumber\\ &&\end{eqnarray}
with the functions $H_{1,3}(m_\pi/T)$ defined by  integrals over the 
Bose distribution of thermal pions:
\begin{eqnarray}
 H_1(y) = \int_y^\infty dx\, {1 \over \sqrt{x^2-y^2}
(e^x-1)}\,,\qquad\qquad H_3(y) = y^{-2} \int_y^\infty dx\, 
{\sqrt{x^2-y^2} \over e^x-1}\,. 
\end{eqnarray}

We proceed with a presentation of results. As input we consistently use the 
same parameters in the chiral limit as in our previous works \cite{cond}, 
namely: $f_\pi = 86.5\,$MeV, $g_A =1.224$, $c_1 =-0.93\,$GeV$^{-1}$ and $M_N = 
882\,$MeV. Concerning the contact term representing unresolved short-distance 
dynamics, we recall from ref. \cite{cond} that its quark mass dependence, 
estimated from recent lattice QCD results \cite{hatsuda}, is negligibly small 
compared to that of the intermediate and long range (pion-exchange driven)
pieces. 

It is worth pointing out again that in-medium chiral perturbation theory with 
this input produces a realistic nuclear equation of state 
\cite{deltamat,FKW2012}, including a proper description of the thermodynamics 
of the liquid-gas phase transition. Apart from temperature $T$, the additional 
``small" parameter in this approach is the nucleon Fermi momentum $p_F$ in 
comparison with the chiral scale, $4\pi f_\pi \sim 1$ GeV. Our three-loop
calculation of the free energy density is reliable up to about twice the 
density of normal nuclear matter.  It can be trusted over a temperature range 
(up to $T\sim 100$ MeV) in which the hot and dense matter still remains well 
inside the phase of spontaneously broken chiral symmetry. 
 
Fig.\ref{figure2} shows a representative example, at  $T = 100$ MeV, displaying 
stepwise the effects of interaction contributions to the density dependence of 
$\langle\bar{q}q\rangle(\rho,T)$ arising from the $m_\pi^2$-derivative 
of the chiral two- and three-body kernels ${\cal K}_{2,3}$. As in the $T=0$ 
case studied previously \cite{cond}, the pion-mass dependence of correlations 
involving virtual $\Delta(1232)$ excitations turns out to be specifically 
important in delaying the tendency towards chiral symmetry restoration as the 
density increases. Once all one- and two-pion exchange processes contributing 
to $\partial {\cal K}_2/\partial m_\pi^2$ and $\partial{\cal K}_3/\partial 
m_\pi^2$ are added up, the chiral condensate at $T=100$ MeV recovers the 
linear density dependence characteristic of a free Fermi gas. However, this 
recovery is the result of a subtle balance between attractive and repulsive 
correlations and their detailed pion-mass dependences. Had we taken into 
account only iterated one-pion and irreducible two-pion exchanges, the system 
would  have become instable not far above normal nuclear matter density as 
can be seen in Fig.\ref{figure2}. In fact, this instability would have 
appeared at even much lower densities in the chiral limit $(m_\pi \rightarrow 
0)$. This emphasizes once more not only the importance of terms including 
$\Delta(1232)$-excitations, but also the significance of
explicit chiral symmetry breaking by small but non-zero quark masses in 
QCD and the resulting physical pion mass, $m_\pi = 135\,$MeV, in governing 
nuclear scales.   

\begin{figure}[htb]
\begin{center}
\includegraphics[width=9.8cm]{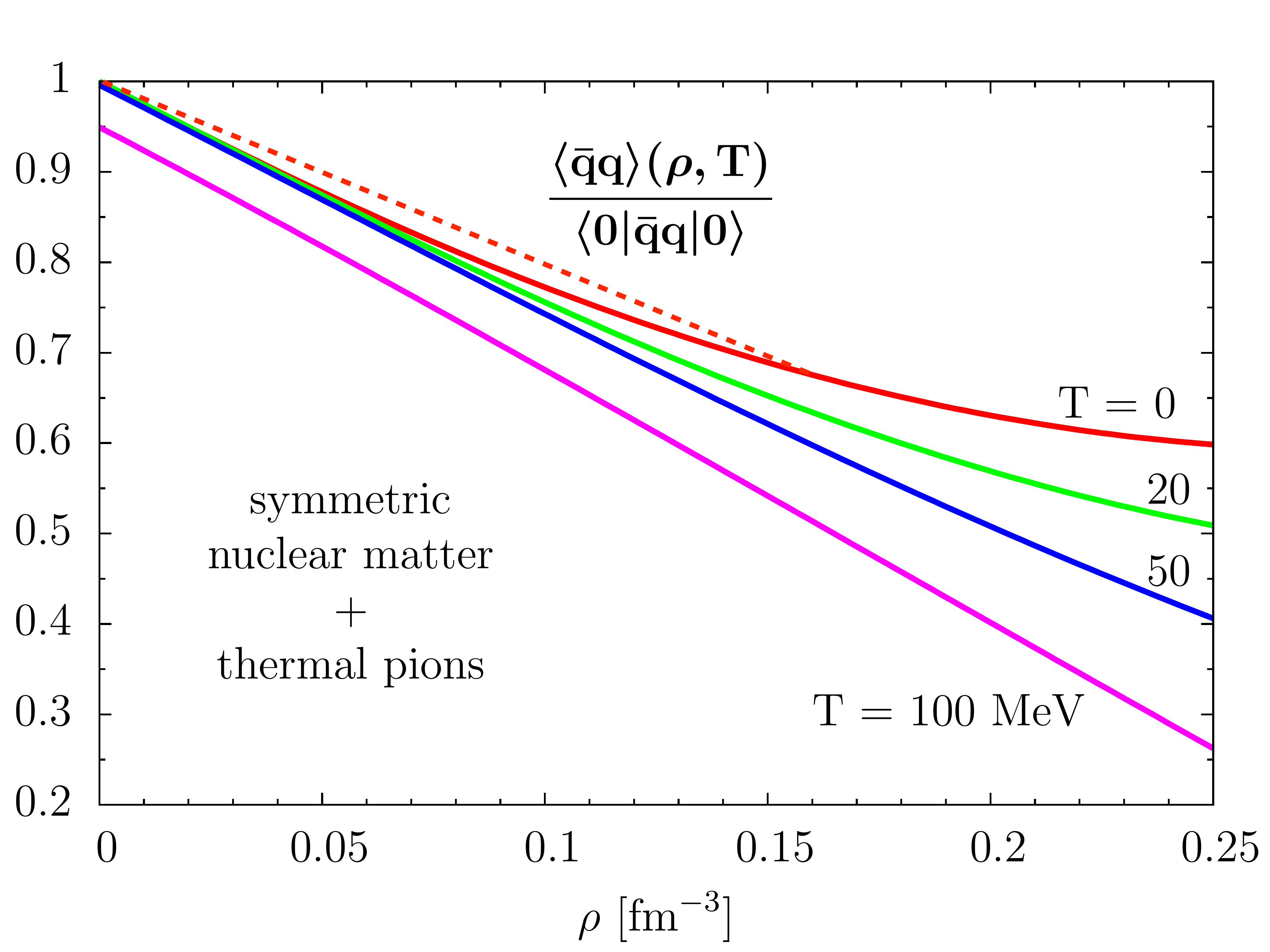}
\end{center}
\caption{Ratio of chiral condensate relative to its vacuum value as function 
of baryon density$\rho$ in isospin-symmetric nuclear matter, for different 
temperatures up to $T = 100$ MeV as indicated. The effects of thermal pions 
are included. The dashed line at $T \simeq 0$ results through a Maxwell 
construction in the coexistence region of the nuclear liquid and gas phases. }
\label{figure3}
\end{figure}

Fig.\ref{figure3} shows the systematics in the variation of the chiral 
condensate with temperature $T$ and baryon density $\rho$. These results 
include all nuclear correlation effects and also the (small) additional shift 
from thermal pions. The latter correction is visible only at the highest 
temperature considered here ($T = 100$ MeV) where the chiral condensate at 
zero density starts to deviate from its vacuum value. The actual crossover 
transition at which the condensate drops continuously to zero is around
$T \sim 170$ MeV \cite{lattice}.  . 

At zero temperature, the hindrance of the dropping condensate at densities 
beyond normal nuclear matter comes primarily from three-body correlations 
through ${\cal K}_3$ which grow rapidly and faster than ${\cal K}_2$ as the 
density increases. The heating of the system reduces the influence of 
${\cal K}_3$ relative to ${\cal K}_2$ continuously as the temperature rises, 
so that their balance at $T=100$ MeV produces a small net effect in 
comparison with the free Fermi gas.  

At $T = 0$, the solid line in Fig.\ref{figure3} does not yet take into 
account the fact that the density range up to and including normal nuclear 
matter density covers the coexistence region of the nuclear liquid and gas 
phases \cite{FKW2012}. Any first-order phase transition is expected 
to leave its mark also in other order parameters, and so it does for the 
chiral quark condensate. Based on the usual Maxwell construction, the dashed 
line in Fig.\ref{figure3} indicates this effect. It becomes much more 
pronounced when the chiral condensate at low temperatures is plotted as a 
function of the baryon chemical potential:
\begin{equation} \mu = M_N +\bigg( 1 + \rho {\partial \over\partial \rho}\bigg) 
\bar F(\rho,T) \,.\end{equation} 

The discontinuity indicating the  first-order liquid-gas transition
at $T$ smaller than the critical temperature for this transition, $T_c \simeq 
15$\,MeV, is clearly visible in Fig.\,4. At this point our results are 
consistent with a recent investigation aimed at an understanding of
chemical freeze-out in heavy-ion collisions at large baryon densities 
\cite{FW2012}, where a similar effect is found using a chiral meson-baryon 
model Lagrangian.  Another side effect induced by the first-order 
liquid-gas phase transition in the discussion of the in-medium chiral 
condensate $\langle \bar q q\rangle(\rho)$ is that the frequently advocated 
``low-density theorem'' needs to be modified:
\begin{equation}  
{\langle \bar q q\rangle(\rho)\over  \langle 0|\bar q q|0
\rangle} = 1 - {\widetilde\sigma_N \over m_\pi^2f_\pi^2} \rho  \,,\end{equation}
where the effective nucleon sigma term $\widetilde\sigma_N\simeq 36\,$MeV 
measures the quark mass dependence of the sum $M_N+\bar E_0$, with $\bar E_0 
\simeq -16\,$MeV the binding energy per particle of saturated nuclear matter. 
The usual version of eq.(16) with the nucleon sigma term $\sigma_N$ in vacuum 
assumes that at sufficiently low densities nuclear matter could be treated as a 
non-interacting Fermi gas.     

\begin{figure}[htb]
\begin{center}
\includegraphics[width=9.8cm]{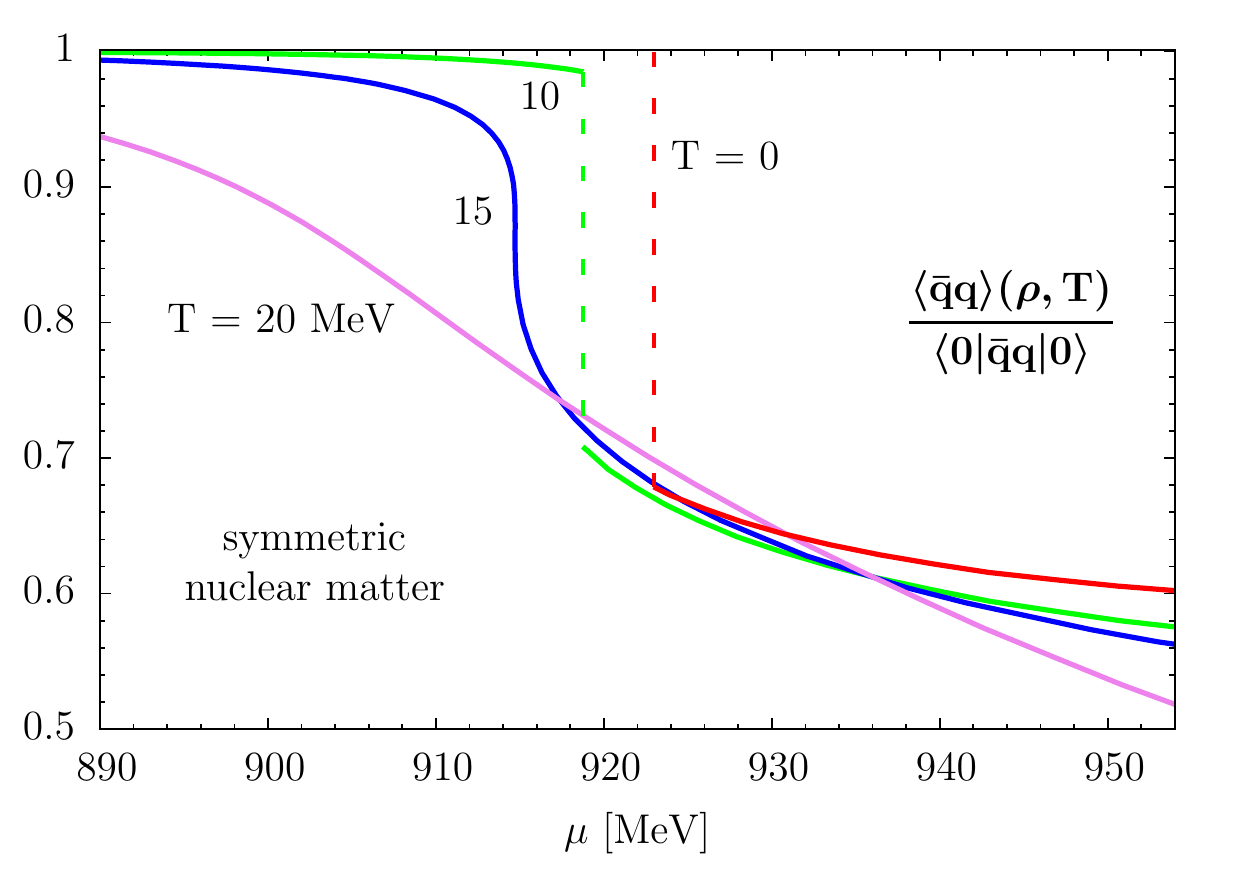}
\end{center}
\caption{Ratio of chiral condensate relative to its vacuum value as function 
of baryon chemical potential in symmetric nuclear matter at low temperatures 
characteristic of the nuclear liquid-gas phase coexistence region.}
\label{figure4}
\end{figure}

In summary, this is the first calculation of the quark condensate at finite 
temperature and density that systematically incorporates chiral two-pion 
exchange interactions in the nuclear medium. Correlations involving 
intermediate $\Delta(1232)$-isobar excitations (i.e. the strong spin-isospin 
polarizability of the nucleon) together with Pauli- blocking effects are 
demonstrated to play a crucial role in stabilizing the condensate at 
densities beyond that of the nuclear matter ground state. The results 
reported here set important nuclear physics constraints for the QCD equation 
of state at baryon densities and temperatures that are of interest e.g. in 
relativistic heavy-ion collisions. In particular, we find no indication of a 
first-order chiral phase transition at temperatures $T\lesssim 100\,$MeV and 
baryon densities at least up to about twice the density of normal nuclear 
matter.  \\

\vspace{1cm}  

{\it Acknowledgements}. This work has been supported in part by BMBF, GSI and 
by the DFG  Cluster of Excellence ``Origin and Structure of the Universe". 
One of us (W.W.) gratefully acknowledges an inspiring
discussion with Christof Wetterich.

\end{document}